--------------------------------------------------------
\documentstyle[prl,aps]{revtex}
\begin{document}
\draft
\title{ Dynamic Domain Walls in Strongly Driven Ferromagnets}
\author{T. Plefka}
\address{ Theoretische Festk\"orperphysik, Technische Hochschule
Darmstadt,
D 64289  Darmstadt, Germany\\
E-Mail: timm@arnold.fkp.physik.th-darmstadt.de}
\date{April 1995}
\maketitle
\begin{abstract}
A multiple-time scaling analysis of
the dissipative, transversely
driven
Landau-Lifshitz equation in presence of exchange, shape
demagnetisation
and week anisotropy fields is performed for a dynamic domain state.
Stationary solutions of the resulting equations
explain  the spatiotemporal structure
of the walls and are
in agreement with previous simulations.
\end{abstract}
\section{Introduction}\label{sec1}
In the interdisciplinary, unifying work \cite{cr93} on  structure
formation in driven dissipative systems ferromagnets are only treated
in exceptional cases. This is surprising as ferromagnets exhibit
already in static fields  domain structures
which are well understood
(for reviews see \cite{dom}) and
which  even serve as textbook examples \cite{textbook}.
Dynamical questions
have  also been investigated. This has been done by extending
the research \cite{dom} to weak driving fields
(compare e.g.\cite{domlin}). For stronger driving fields
numerous work (reviewed in \cite{wigen94})  has been published
to describe the ferromagnetic resonance instabilities \cite {su57}.

Apart from single exceptions like \cite{el88}, where in simulations
dynamic domains for a special model were reported, the existing
investigations on dynamical questions are limited
to the linear or to the weak nonlinear regime.
An approach \cite{Plefka94} to analyse the strongly nonlinear dynamics
of  driven ferromagnets in terms of structure formation has
recently been proposed by the present author. In this approach
a transversely driven model was investigated containing  an exchange,
an anisotropy  and a shape demagnetisation field. Numerical
simulations
exhibited in the rotating frame a stationary domain structure
with  a precessing motion  in the wall regimes.
Some of the characteristic elements of this
structure have been explained in \cite{Plefka94}
by  analytical methods. The spatiotemporal dependence
of the domain walls, however, remained unexplained.

It is the aim of the present investigations to work out analytically
an explanation of these dynamic domain walls. In sec.\ref{sec2}
the basic ingredients of the present approach are given.
After performing a transformation in
sec.\ref{sec3} a multiple time-scaling analysis is employed in
sec.\ref{sec4} to deduce  reduced equations of motion on the slow
time scale. Representing the general result of this work
these eqs. are in sec.\ref{sec5}  applied to determine
the spatiotemporal dependence of the stationary walls. In
sec.\ref{sec6}
the results are compared with the simulations of \cite{Plefka94}.
Finally in sec.\ref{sec7}  some conclusions
are given.

\section{The model and basic assumptions}\label{sec2}

At a mesoscopic scale the dynamics of ferromagnets is governed by the
Landau-Lifshitz equation, which takes the form
\begin{equation} \label{1}
\partial_t\,{\bf m}\,=-(1+ \Gamma {\bf m\times})\;{\bf m\times}\;
(h_\parallel {\bf e}_z \,
 + \, h_\perp{\bf e}_x\,- \,
\bbox{\overline{m}} \,+\, J  \Delta {\bf m}
+A m_z {\bf e}_z)
\,+ \omega {\bf m\times}{\bf e}_z
\end{equation}
in the frame rotating with
the driving frequency
$ \omega $ around the $ {\bf e}_z $ direction.
$ {\bf m}({\bf r},t) $ is the local magnetisation in the
rotating frame being
related to the magnetisation in the laboratory frame $
{\bf m}_{{\rm lab}} $
by $ {\bf m}_{{\rm lab}}\,=\, \exp
(\,  \omega  t {\bf e}_z \times \,)\,\,{\bf m} $.
The amplitudes of the external static and the external
circular driving rf field are denoted by
$ h_\parallel $  and by $ h_\perp $ ,  respectively. The term
$ \bbox{\overline{m}}\,=
\,V^{-1}\int{\bf m}\,{\rm d}V $ represents,  in reduced
units, the demagnetisation field of a sphere of volume $ V $.
The contribution  $ J  \Delta {\bf m}$  results from the isotropic
ferromagnetic exchange interaction and the uniaxial
anisotropy   is described by $ A m_z {\bf e}_z  $ where  $ A<0 $ is
assumed in this work. The Landau-Lifshitz damping
rate is represented by $ \Gamma $.
The gyromagnetic ratio and the magnitude of the magnetisation
$m\,=\, |{\bf m}|$ are set equal to 1.

$ A $ is assumed to be a small quantity  which
permits an perturbative treatment. Recall that eq.(\ref{1}) results
from a continuum approximation in space (compare \cite{Plefka93}) which
implies a slow spatial variation of ${\bf m}$.
Thus the exchange field $ J  \Delta {\bf m}$ is also
treated as a perturbation. Introducing the expansion
parameter $ \epsilon = -A >0 $ and scaling the position as
$ {\bf r}\rightarrow ( -J/A)^{1/2} {\bf r} $
the perturbational fields  of eq.(\ref{1}) $
J  \Delta {\bf m}+A m_z {\bf e}_z$
take the form $ \epsilon ( \Delta {\bf m}- m_z {\bf e}_z) $.

Due to the $ \bbox{\overline{m}}(t) $ term, the problem
described by eq.(\ref{1}) is of the mean field type and the usual
technique
can be applied. This is to employ an
ansatz for $ \bbox{\overline{m}} $, solve formally the problem for
given  $ \bbox{\overline{m}} $ and in the last step examine the results
for self-consistency.

Based of the findings of \cite{Plefka94} as ansatz
a stationary domain structure is assumed  in which the magnetisation
is constant nearly everywhere  and takes only two   values
$ {\bf m_+} $  or $ {\bf m_-} $ realised in the generally
disconnected  partial volumes $ V_+ = n_+V $
and $ V_- = n_-V $ , respectively. The  wall regimes
where  the magnetisation changes from $ {\bf m_-} $ to
$ {\bf m_+} $ are assumed to be narrow. Then
the volume  of the wall regions can be neglected
compared to the sample volume $ V $ which implies
\begin{equation} \label{1a}
{\bf \overline{m}}_{{\rm dom}}=n_+ {\bf m}_+ \;+\;n_- {\bf m}_-
\;;\quad {\rm with}\quad n_+\,+\,n_-\,= \,1\,.
\end{equation}
This work is focusing on the stationary domain solutions of
eq.(\ref{1}) .
To keep the analysis as simple as possible
${\bf \overline{m}}(t)$ is replaced by
$ {\bf \overline{m}}_{{\rm dom}}$ from the beginning.
For a justification of this ansatz it is
pointed out that already for  $ \epsilon = 0 $
relaxation of $ {\bf \overline{m}}(t) $ to
$ {\bf \overline{m}}_{{\rm dom}} $ was found in \cite{Plefka94}.

In this previous work the zeroth order contributions to $ n_\pm $
and to $ {\bf m}_\pm $ have been calculated. The perturbation
theory of this work  makes it
necessary to separate in notation between
full quantities and their contributions to the $ \epsilon $ expansion.
This is archived by generally setting $ a = \tilde{a}\, + \,\epsilon
a^{(1)} + \,{\rm O}(\epsilon^2)\,$ for an arbitrary quantity $ a $.
This rather unconventional notation of the zeroth
order contributions avoids complicated indices.
Employing this notation the results of \cite{Plefka94} take the form
\begin{equation} \label{2}
\tilde{n}_\pm\,=\; \frac{1}{2} \; \pm \frac{d w^2 + \omega u^2}{2 w v}
\quad ; \quad
\tilde{\bf m}_\pm  =\;u^{-1} ( \mp v\,  \tilde{\bf e}_1  \;
+\; \Gamma h_\perp \, \tilde{\bf e}_3  )
\end{equation}
with
\begin{equation}
\label{3}
d  =  h_\parallel -\omega \quad ; \quad
u  =  ( d^2+\Gamma^2)^{1/2} \quad ; \quad
v   = ( u^2-\Gamma^2 h_\perp^2)^{1/2}\; ; \quad
w   =  ( u^2+h_\perp^2)^{1/2}
\end{equation}
and where the transformation from the internal
orthonormal system $ (  \tilde{\bf e}_1 ,  \tilde{\bf e}_2
, \tilde{\bf e}_3 ) $ to
$ ( {\bf e}_x, {\bf e}_y,{\bf e}_z) $ is found to be given by
\begin{equation} \label{4}
 \left(\matrix{{\bf e}_x \cr {\bf e}_y \cr {\bf e}_z} \right)
= \frac{1}{u w} \left(
          \matrix{
                -d h_\perp &         -d u    & \Gamma w \cr
     \Gamma h_\perp & \Gamma u & d w \cr
         -u^2                  &   u h_\perp & 0}
    \right)
 \left(\matrix{
 \tilde{\bf e}_1  \cr \tilde{\bf e}_2  \cr \tilde{\bf e}_3 } \right)\;.
\end{equation}

With eqs.(\ref{2}-\ref{4}) and with the above replacements
eq.(\ref{1}) takes the form (compare \cite{Plefka94})
\begin{equation} \label{6}
\partial_t\,{\bf m}=-\omega w^{-1} \,{\bf m\times}
(\,u\tilde{\bf e}_1 +\, \Gamma h_\perp
{\bf m\times}\tilde{\bf e}_2) -\epsilon\,(1+ \Gamma {\bf m\times})
{\bf m\times}(\Delta{\bf m}-m_z {\bf e}_z-
{\bf \overline{m}}^{(1)}_{\rm dom})\,+{\rm O}(\epsilon^2)
\end{equation}
where $ {\bf \overline{m}}^{(1)}_{{\rm dom}}$ denotes the first order
contributions of $ {\bf \overline{m}}_{{\rm dom}}$  according to
the  convention in notation.

Eq.(\ref{6}) is the starting point for the further analysis.
Compared to eq.(\ref{1}) this eq.(\ref{6}) is constrained by
${\bf \overline{m}}(t)=
{\bf \overline{m}}_{{\rm dom}} $. Note, however, that eq.(\ref{6}) still
describes in general nonlinear deviations from
the domain state as long as these deviations are local.

The   $ {\bf m_\pm} $  have to satisfy eq.(\ref{6})  with
$ \partial_t\,{\bf m}\,=\,\Delta {\bf m}=0 $ from which
\begin{equation} \label{7a}
{\bf m}_\pm^{(1)} =
 \tilde{\bf m}_\pm  {\bf \times}
[\Gamma\,-\, \tilde{\bf m}_\pm  {\bf \times}]\,
[\,\pm \frac{w}{\omega v}{\bf \overline{m}}^{(1)}_{\rm dom}
- \frac{u}{\omega w} \tilde{\bf e}_1
 +\frac{h_\perp}{\omega w} \tilde{\bf e}_2 \,]
\end{equation}
results. Eq.(\ref{1a}) leads in first order to
\begin{equation} \label{7aa}
\overline{\bf m}^{(1)}_{\rm dom}\,=\,n_+^{(1)}\,
(\tilde{\bf m}_+\,-\,\tilde{\bf m}_-)\,+\,
\tilde{n}_+{\bf m}_+^{(1)}\,+\,\tilde{n}_-{\bf m}_-^{(1)}
\end{equation}
where the $ \tilde{\bf m}_\pm $ and the $ \tilde{n}_\pm $ are
given by eq.(\ref{2}). Note that from eqs.(\ref{7a}) and
(\ref{7aa}) the dependence of $ \overline{\bf m}^{(1)}_{\rm dom}\,,
\,{\bf m}_\pm^{(1)} $ and  of $ n_+^{(1)} $ on the model parameters
$ h_\parallel,h_\perp,\omega$ and $ \Gamma $  is not completely
determined.
For such a determination
an additional relation between these quantities is needed. Thus
at this stage of the calculation ${\bf \overline{m}}^{(1)}_{\rm dom} $
and $ {\bf m}_\pm^{(1)} $ are treated
as functions of $ n_+^{(1)} $  and of the model parameters.

\section{Transformations}\label{sec3}

In zeroth order the dynamics  described by eq.(\ref{6})
is hamiltonian \cite{Plefka94}. Thus  the dynamics should be
formulated in variables, which in the $ \epsilon = 0 $ limit reduce
to angle - action variables. The total transformation to these
variables is divided into a sequence of single transformations.
First an  orthonormal system
$ ({\bf e}_1, {\bf e}_2,{\bf e}_3) $ is defined by
$ {\bf e}_1 =(2\,b_1)^{-1}({\bf m}_-\,-\,{\bf m}_+) $ by
$ {\bf e}_3 =(2\,b_3)^{-1}({\bf m}_-\,+\,{\bf m}_+) $ and by
$ {\bf e}_2\,=\,{\bf e}_3 {\bf \times e}_1$
with $ b_1\,=\, \sqrt{ (\,1\,-{\bf m}_+{\bf m}_-)/2} $ and with
$ b_3\,=\, \sqrt{ (\,1\,+{\bf m}_+{\bf m}_-)/2} $ .  As  the
$ {\bf m}_\pm $ are the full domain magnetisations, the
definition of the $ ( {\bf e}_1, {\bf e}_2
,{\bf e}_3) $  is independent of the perturbative treatment
and  holds to all orders in $ \epsilon $.
Note that $\tilde {\bf e}_i $
are the zeroth order contributions to  $  {\bf e}_i $
in accord with the general convention.

In the next steps this $ ( {\bf e}_1, {\bf e}_2
,{\bf e}_3) $ base is first rotated about the (2)-axes
and thereafter standard spherical  coordinates $ \Theta ,\Phi $
with respect to the new (3)-axis are introduced. These
transformations are explicitly described by
\begin{equation} \label{7}
\left(\matrix{{\bf m} \cr {\bf e}_\Theta \cr
{\bf e}_\Phi} \right)=
\left(
\matrix{
\sin \Theta \cos \Phi & \sin \Theta \sin \Phi & \cos \Theta \cr
\cos \Theta \cos \Phi & \cos \Theta \sin \Phi &  -\sin \Theta\cr
-\sin \Phi            & \cos \Phi             &   0  }
    \right)
 \; \left(
\matrix{
b_3\,\cos \Theta &         0   & - b_3 \, p(\Theta)    \cr
               0 &         1   &   0    \cr
b_3 \, p(\Theta)  &       0      &    b_3\,\cos \Theta    }
\right)
 \left(\matrix{
{\bf e}_1 \cr{\bf e}_2 \cr{\bf e}_3} \right)
\end{equation}
where $ p(\Theta)$ is given by
$ p(\Theta)\,=\, \sqrt{b_3^{-2}\,-\,\cos^2 \Theta}$ .
Finally a change of variables from $ \Theta ,\Phi $
to new independent variables $ \Theta ,\varphi $ is performed by setting
\begin{equation} \label{10}
\tan \frac{\Phi(\Theta, \varphi)}{2}= \frac{\Omega}{p(\Theta)- \sin
\Theta}
\tan \Omega \frac{\varphi}{2}\quad {\rm with}\quad
\Omega = \sqrt{b_3^{-2}\,-1}.
\end{equation}
which completes the total transformation.

Eq.(\ref{10}) implies
$ {\rm d} \Phi \,=  p^{-1}(\Theta)\,\cos \Theta
\sin \Phi\,{\rm d} \,\Theta \,+\,g(\Theta, \varphi)\,
{\rm d} \,\varphi $
where the function $ g(\Theta, \varphi) $ is defined as
$ g(\Theta, \varphi)=p(\Theta) + \sin \Theta \cos \Phi $. With
this expression for $ {\rm d} \Phi $ eq.(\ref{7}) leads to
$ \partial_t \,{\bf m} \,=\, g\, p^{-1}\, {\bf e}_\Theta \,\partial_t
\,\Theta \, + \, g \sin \Theta\,{\bf e}_\Phi \,\partial_t \,\varphi $ .
To transform the rhs. of eq.(\ref{6}) in a first step $ \tilde {\bf
e}_1 $
and $ \tilde {\bf e}_2$ are  expressed by the $ {\bf e}_i $ using
eqs.(\ref{2}),(\ref{7a}) and the definition of the $ {\bf e}_i $.
After that eq.(\ref{7}) is employed to represent the rhs. of
eq.(\ref{6})
in terms of $ {\bf e}_\Theta $ and $ {\bf e}_\Phi $.
Introducing  a scaled  time $ \hat{t}$
by $ \hat{t}=\, \omega \,u^2 b_1 b_3 v^{-1}w^{-1}\,t $ and setting
$ \epsilon = \Gamma h_\perp\omega w^{-1} \hat{\epsilon} $
the results of this basically straightforward calculation leads to
\begin{eqnarray} \nonumber
\partial_{\hat{t}}\, \Theta&=&\,\hat{\epsilon} \,
\tilde{p}(\Theta) \,( X_\Phi^{\rm ex} +X_\Phi^{\rm an}+
\Gamma X_\Theta^{\rm ex}+\Gamma X_\Theta^{\rm an}
+Y_\Theta)\,+{\rm O}(\hat{\epsilon}^2)\\ \label{9}
\sin \Theta \,(\partial_{\hat{t}}\, \varphi -1)&=&
\,\hat{\epsilon} \,( -X_\Theta^{\rm ex}-X_\Theta^{\rm an}
+\Gamma X_\Phi^{\rm ex}+\Gamma X_\Phi^{\rm an}+ Y_\Phi)
\,+{\rm O}(\hat{\epsilon}^2).
\end{eqnarray}
The quantities $ \tilde{g}\, X_\Theta^{\rm ex}=
{\bf e}_\Theta \Delta {\bf m}
$ and $ \tilde{g}\, X_\Phi^{\rm ex}={\bf e}_\Phi \Delta {\bf m}  $ are
calculated to
\begin{eqnarray}
\tilde{g}\,X_\Theta^{\rm ex}&=&  \nabla
(\,\tilde{p}^{-1} \tilde{g} \, \nabla \Theta \,)
- \,[\,\tilde{p}^{-1} \sin \Theta \, \nabla \Theta +
\tilde{g}\,\cos \Theta \, \nabla \varphi \,]\,
(\,\tilde{g} \sin \Theta \, \nabla \varphi \,)\nonumber\\
\tilde{g}\,X_\Phi^{\rm ex}&= &\nabla
(\,\tilde{g} \sin \Theta \, \nabla \varphi \,)+
\,[\,\tilde{p}^{-1} \sin \Theta \, \nabla \Theta +
\tilde{g}\,\cos \Theta \, \nabla \varphi \,]\,
(\,\tilde{p}^{-1} \tilde{g} \, \nabla \Theta \,)\,.\label{14}
\end{eqnarray}
{}From the definitions $ \tilde{g}\,X_\Theta^{\rm an}=-({\bf m}
{\bf e}_z)\,({\bf e}_\Theta {\bf e}_z) $ and
$ \tilde{g}\, X_\Phi^{\rm an}=-({\bf m} {\bf e}_z)\,({\bf e}_\Phi
{\bf e}_z) $
\begin{eqnarray} \nonumber
\tilde{g}\,X_\Theta^{\rm an}&=&h_\perp^2 w^{-2}( -\sin \Theta \sin \Phi
+\Gamma \,\tilde{g}\,\cos \Theta)
[\cos \Theta \sin \Phi +\Gamma(\tilde{g} \sin \Theta-\cos \Phi)] \\
\tilde{g}\,X_\Phi^{\rm an}&=&h_\perp^2 w^{-2}( -\sin \Theta
\sin \Phi +\Gamma \,\tilde{g}\,\cos \Theta)
( \cos \Phi+\Gamma \cos \Theta  \sin \Phi)\label{15}
\end{eqnarray}
is obtained and for   $ Y_\Theta $ and $ Y_\Phi $
\begin{eqnarray}
Y_\Theta\,&=& -\Gamma^2 h_\perp v^{-2}
[\,u\,\tilde{\bf e}_1 {\bf \overline{m}}^{(1)}_{\rm dom}
\,+\,h_\perp\,\tilde{\bf e}_2{\bf \overline{m}}^{(1)}_{\rm dom}  ]
\,\sin \Theta\,\nonumber \\
& &-\,\Gamma h_\perp^2 w^{-2}\tilde{g}^{-1}
[\,(1-\Gamma^2) \cos \Theta\,( \tilde{p} \cos \Phi\,+ \sin \Theta)\,+
\, 2 \Gamma \tilde{p} \sin \Phi]\nonumber\\
Y_\Phi\,&=& \Gamma h_\perp v^{-2} u^{-1}
[\,(v^2-\Gamma^2 h_\perp^2)\,
\tilde{\bf e}_1{\bf \overline{m}}^{(1)}_{\rm dom}  \,+
\,\Gamma^2 h_\perp u \,
\tilde{\bf e}_2{\bf \overline{m}}^{(1)}_{\rm dom}  ] \,\sin
\Theta\,\nonumber \\
& & +
\,\Gamma h_\perp^2 w^{-2}\tilde{g}^{-1}
[\,-2 \Gamma \cos \Theta\,( \tilde{p} \cos \Phi\,+ \sin \Theta)\,+
\, (1- \Gamma^2) \tilde{p} \sin \Phi]\label{16}
\end{eqnarray}
results with $ \Phi = \Phi(\Theta, \varphi)$ as defined by
eq.(\ref{10}). The quantities $  \tilde{p} $ and $ \tilde{g} $
stand for the zeroth order contributions to $ p $ and to $ g $
and are explicitly calculated to
\begin{equation}
\tilde{p}(\Theta)=\sqrt{\tilde{\Omega}^2\,+\, \sin^2 \Theta}\quad,
\quad
\tilde{g}(\Theta, \varphi)=\tilde{p}(\Theta) + \sin \Theta \cos \Phi
\quad {\rm with} \quad \tilde{\Omega}=\, \frac{v}{\Gamma h_\perp}.
\end{equation}

In the zeroth order limit  eqs.(\ref{9})
reduce to $ \partial_{\hat{t}}\, \Theta\,=\,0 $ and to
$ \partial_{\hat{t}} \,\varphi\,=\,1 $ . Thus as required
the variables $ \Theta ,\varphi $ become
angle - action variables in the $ \epsilon=0 $ limit.
A second  feature of the transformation should be pointed
out. According to the definition of the $ {\bf e}_i $ and to
eq.(\ref{7})
$ {\bf m}(\Theta \rightarrow 0,\varphi)={\bf m}_- $ and
$ {\bf m}(\Theta \rightarrow \pi,\varphi)={\bf m}_+ $ holds.
Thus the values $ \Theta=0 $ and  $ \Theta=\pi $
characterise the interior of the domain regimes
and a domain  wall corresponds to  a change from $\Theta=0$ to
$\Theta=\pi$.
Note that, as it must be, eqs.(\ref{9})   are satisfied
for  $ \Theta=0 $ and  for $ \Theta=\pi $ with $ \partial_{\hat t}\,
\Theta\,=\,\nabla\Theta=0 $.

\section{Multiple-Time Scaling}\label{sec4}
Next the standard multiple-time scaling method is applied
to the eqs.(\ref{9}) . Denoting the slow time scale
as  $ T=\hat{\epsilon}\,\hat{t} $  the ansatz
\begin{eqnarray} \nonumber
\Theta(\hat{t},{\bf r})&=& \tilde{\Theta}(T,{\bf r})+
\hat{\epsilon}\, \Theta^{(1)}(\hat{t},T,{\bf r}) \,+\,
{\rm O}(\hat{\epsilon}^2) \\
\varphi(\hat{t},{\bf r})&=& \hat{t}  +\tilde{\varphi}(T,{\bf r})+
\hat{\epsilon} \varphi^{(1)}(\hat{t},T,{\bf r})+\,
{\rm O}(\hat{\epsilon}^2)\label{19}
\end{eqnarray}
is employed which already satisfies the zeroth order
of the perturbative expansion.
In first order   the $ \hat{t} $-secular terms of
$ \Theta^{(1)} $ and of $ \varphi^{(1)} $ are as usual eliminated.
This
procedure  leads to
\begin{eqnarray} \nonumber
\partial_{T}\, \tilde{\Theta}&= &\tilde{p}( \tilde{\Theta})\,
( \,\langle X_\Phi^{\rm ex} \rangle +\,\langle X_\Phi^{\rm an} \rangle
 \,+\,\Gamma \langle X_\Theta^{\rm ex} \rangle \,
+\,\Gamma \langle X_\Theta^{\rm an} \rangle \,
+\,\langle Y_\Theta \rangle \,) \\
\label{25}
\sin \tilde{\Theta} \;\partial_{T} \,\tilde{\varphi}&=&
-\,\langle X_\Theta^{\rm ex} \rangle \,
-\,\langle X_\Theta^{\rm an} \rangle \,
\,+\,\Gamma \langle X_\Phi^{\rm ex} \rangle
\,+\,\Gamma \langle X_\Phi^{\rm an} \rangle
+\,\langle Y_\Phi \rangle
\end{eqnarray}
which determines the $ T $- dependence of
$ \tilde{\Theta}(T,{\bf r})$ and of $\tilde{\varphi}(T,{\bf r}) $.
In eqs.(\ref{25})
$\langle \,f(\Theta,\varphi) \,\rangle
=\tilde{\Omega}/(2 \pi)  \int^{2 \pi\tilde{\Omega}^{-1}}_0
\,f(\tilde{\Theta},\hat{t}+\tilde{\varphi})\,{\rm d} \hat{t} $
denotes an averaging of an arbitrary function
$ f(\Theta,\varphi)$ over the fast time variable $ \hat{t} $.
Using a substitution analogous to eq.(\ref{10}) these averages
can be performed and result in
\begin{eqnarray} \label{16a}
\langle X_\Theta^{\rm ex} \rangle &=&
(\tilde{p} +\tilde{\Omega})\nabla \, \frac{\nabla  \,\tilde{\Theta}}
{\tilde{p}(\tilde{p} +\tilde{\Omega})}
\,- \tilde{\Omega} \sin \tilde{\Theta}\,
\cos \tilde{\Theta} \,( \nabla  \,\tilde{\varphi})^2\nonumber\\
\langle X_\Phi^{\rm ex} \rangle&=&
\,\sin \tilde{\Theta}\,\frac{ \tilde{p} +\tilde{\Omega}}
{\tilde{p} -\tilde{\Omega}}\,\nabla\,(\frac{ \tilde{p} -\tilde{\Omega}}
{\tilde{p} +\tilde{\Omega}}\,\nabla  \,\tilde{\varphi})
\nonumber\\
\langle X_\Theta^{\rm an} \rangle &=&
h_\perp^2 w^{-2} \,(\tilde{\Omega} - \tilde{p})\,[ 1\,-\Gamma^2(
1\,+\tilde{\Omega} \tilde{p}+
\tilde{\Omega}^2)]\; \cot \tilde{\Theta}
\nonumber\\
\langle X_\Phi^{\rm an} \rangle&=&\,2\,\Gamma\,
h_\perp^2 w^{-2} \,(\tilde{\Omega} - \tilde{p})\, \cot \tilde{\Theta}
\nonumber \\
\langle Y_\Theta\rangle\,&=& -\Gamma^2 h_\perp v^{-2}
[\,u\,\tilde{\bf e}_1 {\bf \overline{m}}^{(1)}_{\rm dom}
\,+\,h_\perp\,\tilde{\bf e}_2{\bf \overline{m}}^{(1)}_{\rm dom}  \,]
\,\sin \tilde{\Theta}\,\nonumber \\
\langle Y_\Phi \rangle\,&=& \Gamma h_\perp v^{-2} u^{-1}
[\,(v^2-\Gamma^2 h_\perp^2)\,\tilde{\bf e}_1 {\bf
\overline{m}}^{(1)}_{\rm dom} \,+
\,\Gamma^2 h_\perp u \,
\tilde{\bf e}_2 {\bf \overline{m}}^{(1)}_{\rm dom} \,] \,\sin
\tilde{\Theta}.
\end{eqnarray}
where now $\tilde{p}$ stands for
$\tilde{p}(\tilde{\Theta})$.

Together with the expressions (\ref{16})
the coupled  set of eqs.(\ref{25})
represents the general result of this work. On the slow time scale $ T $
the temporal and the spatial evolution of the magnetisation is
governed by these reduced equations of motion. It is expected  that
various, interesting questions can be  analysed on the basis
of these equations. One of these problems, the profile
of a stationary  domain wall is treated in the next section.

\section{Domain Wall}\label{sec5}
In this first approach to the domain wall problem I restrict myself
to planar walls whose spatial dependence varies -
locally and/or approximately - only in one direction,
the normal direction $ {\bf e}_\xi $ of the wall.
The analysis is further restricted
to stationary walls. Thus   $ \tilde{\Theta}(\xi)$
is assumed  to be  time independent.  For the cyclic variable
$ \tilde{\varphi} $ the ansatz $\tilde{\varphi}(T,\xi)=
\,\Omega^{(1)}\,T \,
\,+\,\bar{\varphi}({\xi}) $ is used where $ \Omega^{(1)} $ describes
a first order frequency contribution independent of $\xi$.
Fixing the scale of $ \xi $ by the replacement
$ ({\bf r}{\bf e}_\xi)  \rightarrow  h_\perp w^{-1} \xi$
and introducing the ' local wave number variable ' by
\begin{equation}\label{32}
Q(\xi)=\, \tilde{\varphi}^\prime (T,\xi)
\end{equation}
eqs.(\ref{25}) lead to a system
of ordinary differential eqs. for $\tilde{\Theta(\xi)} $ and for
$ Q(\xi)$
\begin{eqnarray}\nonumber
(\tilde{p} +\tilde{\Omega})\biggl(
\frac{\tilde{\Theta}^\prime }
{\tilde{p}(\tilde{p} +\tilde{\Omega})}\biggr)^\prime
\,- \tilde{\Omega} \sin \tilde{\Theta}\,
\cos \tilde{\Theta} \,Q^2\,&=&\,( \tilde{p}-\tilde{\Omega})\,[
1\,-\Gamma^2( 1\,+\tilde{\Omega} \tilde{p}+
\tilde{\Omega}^2)]\; \cot \tilde{\Theta}\,+
\, C_\Theta \,\sin \tilde{\Theta}\\
\sin \tilde{\Theta}\,\frac{ \tilde{p} +\tilde{\Omega}}
{\tilde{p} -\tilde{\Omega}}\,\biggl(\frac{ \tilde{p} -\tilde{\Omega}}
{\tilde{p} +\tilde{\Omega}}\,Q\biggr)^\prime\,&=&\,2\,\Gamma\,
(\tilde{p}-\tilde{\Omega})\, \cot \tilde{\Theta}\,+
\, C_\Phi \,\sin \tilde{\Theta}
\label{33}
\end{eqnarray}
where  $\tilde{p}$ still stands for
$\tilde{p}(\tilde{\Theta})=\sqrt{\tilde{\Omega}^2\,+\,
\sin^2 \tilde{\Theta}}$ and
where the constants $ C_\Theta $ and $ C_\Phi $ satisfy the linear
system
\begin{eqnarray} \label{35}
\Gamma C_\Theta \,+\, C_\Phi \,&=&\,
\Gamma^2 w^2 h_\perp^{-1} v^{-2}
[\,u\,\tilde{\bf e}_1{\bf \overline{m}}^{(1)}_{\rm dom}
\,+\,h_\perp\,\tilde{\bf e}_2{\bf \overline{m}}^{(1)}_{\rm dom}  \,]\\
\label{35a}
- C_\Theta \,+\, \Gamma C_\Phi \,&=&\,w^2 h_\perp^{-2}\, \Omega^{(1)}
\,-\,\Gamma w^2 h_\perp^{-1} v^{-2} u^{-1}
[\,(v^2-\Gamma^2 h_\perp^2)\,\tilde{\bf e}_1
{\bf \overline{m}}^{(1)}_{\rm dom}\,+
\,\Gamma^2 h_\perp u \,
\tilde{\bf e}_2{\bf \overline{m}}^{(1)}_{\rm dom}  \,].
\end{eqnarray}

Apart from solving eqs.(\ref{33})  $ \tilde{\Theta}(\xi) $
and $ Q(\xi)$ have to satisfy at the boundaries
\begin{equation} \label{37}
\tilde{\Theta}( \xi \rightarrow -\infty)=0 \,;\quad
\tilde{\Theta}( \xi \rightarrow \infty)=\pi \,;\quad
\tilde{\Theta}^\prime( \xi \rightarrow \pm \infty)=0 \,;\quad
Q( \xi \rightarrow \pm \infty)=const.
\end{equation}
where  the range of $ \xi $ has been extended
over the entire interval from $  -\infty $  to $ \infty $.
Just by counting constants of integration it is
obvious that the  conditions (\ref{37}) can  generally
not be satisfied, noting additionally  that one constant
is the center of the wall $ \xi_0 $ due to translation invariance of
eqs.(\ref{33}). For special values of $ C_\Theta $
and $ C_\Phi $, however, solutions of this  boundary eigenvalue
problem may exist.

For $ C_\Theta = C_\Phi =0 $ a solution exists which exhibits the
symmetry properties $ \tilde{\Theta}(\xi)\,=
\,\pi -\tilde{\Theta}(-\xi) $ and $ Q(\xi)=Q(-\xi)$. This will be
demonstrated by construction and for this propose
eqs.(\ref{33}) are formally integrated
\begin{eqnarray} \nonumber
(\tilde{\Theta}^\prime)^2&=&\,2
\tilde{p}^2\,(\tilde{p}+\tilde{\Omega})^2
\int_{\tilde{\Omega}}^{\tilde{p}}\;{\rm d}p
\left \{
\frac{1-\Gamma^2 (1+p\tilde{\Omega}+
\tilde{\Omega}^2)}{(p+\tilde{\Omega})^3}\,+\,\frac{\tilde{\Omega}
\,Q^2}{(p+\tilde{\Omega})^2} \right\}
\\ \label{36}
Q&=&\,2\,\Gamma\, \frac{\tilde{p}+\tilde{\Omega}}{\tilde{p}-
\tilde{\Omega}}
\int_{\tilde{\Omega}}^{\tilde{p}}\;{\rm d}p\;
\frac{p(p-\tilde{\Omega})^{1/2}}{(p+\tilde{\Omega})^{5/2}
\,\tilde{\Theta}^\prime}.
\end{eqnarray}
Consider both $ \tilde{\Theta}^\prime $ and $ Q $ as functions of
$ \tilde{\Theta} $ or according to the relation $ \sin \tilde{\Theta}
=( \tilde{\Omega}^2-\tilde{p}^2)^{1/2} $ as functions of $ \tilde{p} $.
Then the eqs.(\ref{36}) represent a system of integral eqs. for
$ \tilde{\Theta}^\prime $ and $ Q $ in dependence of $ \tilde{p} $.
Denoting  the explicit solutions of this system
by $ F(\tilde{p})= \tilde{\Theta}^\prime $ and by $ G(\tilde{p})= Q $
further integration leads to $
\xi\,= \int_{\pi/2}^{\tilde{\Theta}(\xi)}
F^{-1}( \sqrt{\tilde{\Omega}^2+\sin^2 \Theta })\,{\rm d} \Theta $
where the center of the wall has been fixed by $
\tilde{\Theta}(0)=\pi/2 $.
Thus the $\xi$ dependence  of ${\tilde{\Theta}(\xi)}$ is implicitly
determined  from which $ Q(\xi) $ and
consequently  $ \tilde{\varphi}(T,\xi) $ can be calculated according to
eq.(\ref{32}).

In the limit $ \tilde{p} \rightarrow \tilde{\Omega} $ which
corresponds to the limit $ \xi \rightarrow \pm \infty $ the
eqs.(\ref{36})
lead to $ F(\tilde{p})\sim (\tilde{p} -\tilde{\Omega})^{1/2}
\rightarrow 0 $ and to $ G(\tilde{p})\rightarrow $ const. .
These findings are valid for all parameter values of $\tilde{\Omega} $
and of $\Gamma$ and imply that all the boundary conditions
(\ref{37}) are satisfied.

Turning now to the  question whether further solutions of the boundary
eigenvalue problem exist eqs.(\ref{33}) are again integrated formally.
Due to $ C_\Theta \ne 0 $ and $ C_\Phi \ne 0 $ additional contributions
arise to eqs.(\ref{36}). From these contributions
the asymptotic behaviour for $ \xi\rightarrow \infty $
$ \bigl(\tilde{\Theta}^\prime(-\xi)\bigr)^2 -
\bigl(\tilde{\Theta}^\prime(\xi)\bigr)^2 \,\sim C_\Theta $
and $ Q(-\xi)-Q(\xi)\,\sim
\,C_\Phi\,(\tilde{p}- \tilde{\Omega})^{-1/2}$
can be deduced. These results
imply that the conditions (\ref{37}) can not be satisfied
with  $ C_\Theta \ne 0 $ or with $ C_\Phi \ne 0 $. Thus there are
no further solutions and  $ C_\Theta = C_\Phi = 0 $ represents
a compatibility condition.

In general the explicit solutions of
eqs.(\ref{33}) with $ C_\Theta = C_\Phi = 0 $
have to be calculated numerically.
Fig.(1) shows the results of such a calculation for the parameter
values
$ \tilde{\Omega}=\sqrt{3} $ and $ \Gamma =0.1 $ . Note that
the values of $ Q $  are small compared to the variation of
$ \tilde{\theta} $. This findings imply that the spatial dependence
of the wall is mainly determined by $ \tilde{\Theta}(\xi) $
and are  attributed to the fact that
$ Q(\xi) $ is proportional to $ \Gamma $. Thus alternatively
an iteration procedure can be employed to find
approximate solutions. Starting with $ Q_0(\xi)= 0 $  at the rhs.
of eqs.(\ref{36}) these eqs. determine first approximations.
For the above parameter values the deviations from the exact
solutions are found to be small.
As usually $ \Gamma \ll 1 $ is  satisfied this
indicates that the first iteration of eqs.(\ref{36}) - which
can be calculated analytically - may be sufficient in many cases.

According to eq.(\ref{35}) $ C_\Theta = C_\Phi = 0 $  implies
the relation $ u\,\tilde{\bf e}_1{\bf \overline{m}}^{(1)}_{\rm dom}
\,+\,h_\perp\,\tilde{\bf e}_2{\bf \overline{m}}^{(1)}_{\rm dom}=0  $.
This additional relation to the eqs.(\ref{7a}) and (\ref{7aa})
now  determines in a unique way the dependence
of $ n_+^{(1)}$ of $ {\bf m}_\pm^{(1)} $ and of $
{\bf \overline{m}}^{(1)}_{\rm dom} $ on the model parameter.
With eq.(\ref{35a}) this also applies to $ \Omega^{(1)} $.
Thus all quantities of the domain states including the
spatiotemporal wall structure are uniquely determined.
This result together with the fact that the wall width
remains finite in the large $ V $ limit shows the self-consistency of
the ansatz(\ref{1a}).

A further consequence of the additional relation is
$\langle Y_\Theta \rangle \,=0 $ and thus this term in
eqs.(\ref{25}) drops out. By a redefinition of the fast time scale
in the first order term, it is possible to transfer
the contribution  $\langle Y_\Phi \rangle  $ of  eqs.(\ref{25})
to the fast dynamics. Employing this redefinition both
$\langle Y_\Theta \rangle \, $ as well as
$\langle Y_\Phi \rangle  $ drop out  of eqs.(\ref{25}).
This findings considerably simplify the analysis of further
problems based on the general eqs.(\ref{25}).

Next the internal fields ${\bf H}_1 $ and $ {\bf H}_2 $
are introduced by $ {\bf H}_2={\bf H}_1+ \omega {\bf e}_z
=h_\parallel {\bf e}_z \,
 + \, h_\perp{\bf e}_x\,- \,
\bbox{\overline{m}}_{\rm dom} $.  Employing again
the above relation and eqs.(\ref{2}-\ref{4}) it is found
that up to the first order
\begin{equation}\label{40}
{\bf H}_1 \, {\bf H}_2 =0
\end{equation}
holds which generalises the zeroth order result of \cite{Plefka94}.
In this ref. the importance of eq.(\ref{40}) as a
criterium for the existence of the domain states has already been
discussed.

To conclude this section it is mentioned that within the domain regions
a linear  stability analysis  has been performed with the
result that all the in zeroth order undamped modes of \cite{Plefka94}
have now finite damping constants proportional
to $ \epsilon $. This stability
analysis does not include the wall regimes as for such a complete
treatment numerical methods have to be applied. The complete zeroth
order stability analysis, the present partial first order analysis
and the results of the computer simulations yield considerable
evidence for
the stability of the dynamic domain states. In this context it is
pointed out that in the weak coupling limit $ \Gamma\ll 1$ a
Lyapunov
function can be constructed on the slow time scale which implies
stability.

\section{Comparison with Simulations}\label{sec6}

With the numerical results of the last section and an additional
integration the zeroth order contributions of eqs.(\ref{19}) are
determined. With these results  for  $ \Theta $  and $ \varphi $
the Cartesian components of $  m_i $ can be calculated
according to eq.(\ref{7}). This has been done for the parameter
values $ h_\parallel=\omega=2,\; h_\perp=0.5 $ and $ \Gamma=0.1 $
neglecting $ \Omega^{(1)}$  and all  other first order
corrections in eq.(\ref{7}) which  implies also
$ {\bf e}_i =\tilde{\bf e}_i $. The results of this calculation
are given in Fig.(2a) and Fig.(2b) where the spatiotemporal
dependence of $ \cos \phi= m_1 (m_1^2+ m_2^2)^{-1/2} $ and of $ m_3 $
are presented. For the time dependence  anharmonic oscillations
are found which are plotted   on the $\tau= \tilde
{\Omega} \hat{t}$ scale for one period.

Fig.(2c) and Fig.(2d) show the corresponding results \cite{Plefka94}
of the computer simulations
of eq.(\ref{1})  performed with the same parameter values and using
in addition $ J=0.01$ and $ A=-0.005 $. The original simulation
results are
rescaled to the theoretical scales. With the exception of the center
of the wall and the time origin no fitting procedure
has been used.

With deviations of usually less than one percent
the analytic results agree with the simulations.
Quantitative agreement for $ {\bf \overline{m}}_{\rm dom} $, for
$ {\bf m}_\pm  $ and for $n_+$ was already found in \cite{Plefka94}
with a similar high accuracy. Based on these agreements
it has been demonstrated that the employed methods
are a suitable tool to explain the
characteristic features of the dynamic domain states.

\section{Conclusions}\label{sec7}

In this work a rather realistic model for a ferromagnet was
investigated. Depending on the parameter values this model
can exhibit  stationary dynamic domain states in the rotating frame.
An important element of this states is the spatiotemporal
structure of the domain walls. With the present investigation
this structure is understood to a high extent.

The analysis of wall structure is based on the reduced
equations of motion (\ref{25}) which  from the theoretical
point of view represent the main result of this work.
These eqs. which govern the temporal and the spatial
evolution on the slow time scale should be applicable to
other phenomena. Obvious extensions are investigations on drifting
walls, on multi-dimensional walls  and on the interaction
between domain walls.

As a further propose the present approach should be extended
to take stray fields into account.  This  would imply that the
dipole interaction
is completely included and the model will become
very realistic for ferromagnetic materials. Moreover by analogy
with the static case, it is expected that these dipolar stray fields
contribute crucially to selection mechanism for the wall positions.
In this context the formation of regular patterns in driven
ferromagnets
seems to be possible.

The author has benefited from discussions with
W. Just and G. Sauermann. This work was performed within SFB 185.

\begin{figure}
\caption{ Numerical solution of eq.(20) (with $ C_\Theta=C_\Phi=0 $ )
for $ \tilde{\Theta}(\xi) $ and $ Q(\xi) $ satisfying
the boundary conditions. of eq.(23).
The parameter values are
$ \tilde\Omega= 3^{1/2}$
and $ \Gamma =.1 $.
}
\label{Fig.1}
\end{figure}

\begin{figure}
\caption{
Spatiotemporal dependence of a stationary  domain wall in
reduced units $ \xi$ and $\tau$. In Fig.(2a) and (2b)  the
results of the analytical treatment of this work are presented.
For comparison  Fig.(2c) and (2d) show the results of the numerical
simulations of [8].
For both cases  $ \cos \phi = m_1 ( m_1^2 +m_2^2 ) ^{-1/2}$
and $m_3$ are plotted, where $ m_i $ are the components
of the local magnetisation in the internal coordinates, defined by
eqs.(4) and (5). The scaled position
$ \xi $  and scaled time $ \tau $ are given by
$ \xi=w h_\perp^{-1}(-A/J)^{1/2}\, {\bf e}_\xi{\bf r}$
and by $ \tau= \tilde{\Omega}\,\hat{t}=\omega v w^{-1} \,t$,
respectively. Compare text for the parameter values.
\label{Fig.2}}
\end{figure}
\end{document}